# Exemplary and Complete Object Interaction Descriptions*

Ruth Breu, Radu Grosu, Christoph Hofmann,
Franz Huber, Ingolf Krüger, Bernhard Rumpe,
Monika Schmidt, Wolfgang Schwerin

email: {breur,grosu,hofmannc,huberf,kruegeri,rumpe,schmidtm,schwerin}
@informatik.tu-muenchen.de

Technische Universität München
Arcisstr. 21
D-80290 München, Germany

In this paper, we present a variant of message sequence diagrams called EETs (Extended Event Traces). We provide the graphical notation, discuss the methodological use of EETs to describe behavior of object-oriented business information systems, and sketch their semantics. Special emphasis is put on the different implications of using EETs for exemplary and complete interaction descriptions.

The possibility to describe interactions between single objects as well as composite objects with EETs makes them particularly suitable to describe the behavior of large systems.

## 1   Introduction and Classification

Message sequence diagrams are one of the most widely accepted graphical techniques for describing the dynamic behavior of systems. Originally developed for the design of technical systems (in particular: telecommunication protocols, cf. [IT94, IT96]), sequence diagrams have recently become increasingly popular in the field of business information systems. Most object-oriented analysis and design techniques offer some notational variant of sequence diagrams to express object interaction [RBP+91, Boo94, J+92, BRJ96].

---

*This paper is joint work of the members of the projects SYSLAB (supported by the DFG under the Leibniz Program, and by Siemens-Nixdorf), Arcus (supported by the BMBF under project name 'ENTSTAND'), and Bellevue (supported by the DFG).

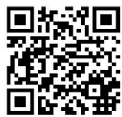

Although most developers consider sequence diagrams intuitive and easy to draw, the exact interpretation and systematic use of these diagrams is often neglected. In this paper we both discuss the methodological use of sequence diagrams within the design process for information systems and give an overview of how sequence diagrams can be assigned an exact, formal semantics.

The adequate semantic interpretation of sequence diagrams depends on various factors. One basic issue is the underlying mechanism for message exchange (synchronous vs. asynchronous). While for telecommunication protocols the ability to model asynchronous message transmission is of the essence, synchronous communication might be a more adequate model for certain classes of business information systems.

Another fundamental issue (especially in the context of object-oriented modeling) addresses the question how sequence diagrams determine the life cycles of objects. In order to answer this question, the role of sequence diagrams within the design process has to be understood. We identify two substantially different ways how sequence diagrams can be used.

One possibility is to use sequence diagrams for describing exemplary behavior of objects (*scenarios*). Scenario-based design of systems has evolved as an excellent paradigm both to explore the requirements of a system and to communicate with customers in an appropriate way. Here, a sequence diagram describes a scenario as a single trace of events, i.e. trace of messages exchanged by the objects of the system under design.

Alternatively, sequence diagrams can also be used to describe complete object behavior, i.e. the set of all possible interaction sequences during the lifetime of the modeled system components. It is clear that the basic form of sequence diagrams (providing only simple syntactical elements for message exchange) is not powerful enough to describe complete object behavior, and a set of additional operators has to be defined. Repetition, alternative, and hierarchical structuring operators are of particular importance. Among more elaborate extensions are recursion or feed-back constructs and conditions on object states. Notations that provide some or all of these operators are MSC96 [IT96], EETs [BHKS97, SHB96], and also most of the sequence diagrams supported by the OOA methods mentioned above.

Our claim is that these two views of sequence diagrams require a different semantic interpretation and address different design issues.

Scenarios describe single sequences of messages. Their semantics has to cope with the question how exemplary object behavior can be interpreted in the context of complete object behavior. This question is of importance when combining different description techniques like sequence diagrams and state transition diagrams. Many object-oriented analysis methods offer both of these description techniques but do not provide adequate guidelines, let alone a semantic foundation, for their consistency and combination.

In contrast, sequence diagrams used for a complete description of object behavior impose conditions on the overall set of messages that an object can send or receive. Since the notations mentioned earlier support complex constructs, such as repetition and alternative, the semantic foundation of these constructs helps designers to obtain a better understanding of the diagrams they develop. Other important issues are the combination of different diagrams involving the same set of objects and the refinement of diagrams during design.

Summarizing, the main aim of this paper is to discuss the semantical and methodological questions that arise when sequence diagrams are applied in the design process for object-oriented business information systems.

For the semantic model we use our experience gained while working on the semantic foundation of software engineering techniques and software architecture using formal methods. In particular, our group has developed a formal model, called *system model* (cf. [KRB96]), which is capable of modeling various system views that arise during the development process. These views are modeled abstractly and independently of specific description techniques.

The structure of this paper is as follows. In Section 2 we provide examples for the application of sequence diagrams as a description technique for exemplary (Section 2.1) and complete object interactions (Section 2.2). In Section 3 we sketch the semantic foundation. We discuss our conclusions in Section 4.

## 2  Notation and Methodology

The basic constituents of sequence diagrams are components and events. In general, a component may be a single object or a composite object performing a certain task. Henceforth, we use the notion of component and object interchangeably.

Each object that participates in an interaction is depicted by a vertical axis (labeled with the object name) that represents (part of) the lifetime of the object where time advances from top to bottom. An interaction is indicated by an arrow directed from the sending object to the receiver. Interactions are labeled with a message name and an optional list of parameters.

Figure 1 depicts a first view of a car rental company, an example which we will employ throughout this paper. The given sequence diagram models a customer making a car reservation at some reservation branch.

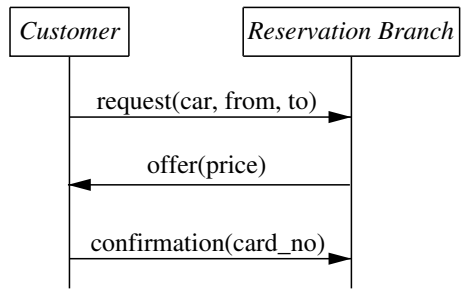

Figure 1: Simplified View of a Car Reservation

The diagram of Figure 1 can be interpreted in a number of different ways:

- Customer and reservation branch are able to exchange and to react to the given sequence of messages.

- Every time the customer sends message *request* to the reservation branch, the messages *offer* and *confirmation* have to be exchanged consecutively (with the variation that customer and reservation branch may or may not send or receive other messages in between).

- The diagram describes the full interaction between the customer and the reservation branch, i.e. both objects can only interact observing the protocol defined by the sequence of messages given in the diagram.

The above interpretations differ in the degree in which they constrain the overall behavior of the objects involved. Most analysis methods leave open how sequence diagrams are interpreted in their framework. Since an analysis model is a kind of contract either between the customer and the developer or the designer and the implementor, severe misunderstandings may be the consequence.

On the other hand, our observation is that there is no "best" interpretation of sequence diagrams and their semantics is heavily influenced by the context in which they are used. For this reason, the rest of this section deals with the methodological aspects of design with sequence diagrams. Section 2.1 focuses on sequence diagrams describing exemplary behavior of objects while Section 2.2 is concerned with the description of complete interaction behavior.

## 2.1 Exemplary Descriptions

In Figure 2 an interaction pattern of a successful car reservation is depicted. Interpreting the given sequence diagram as an exemplary scenario, i.e. as one particular sequence of interactions possibly occuring in the system under specification, the following system properties are expressed:

- Structurally, the system encompasses at least one reservation branch component, one pickup branch component, and one customer component.

- With respect to the system's behavior the following must be true: Upon receiving a *request* message with three formal parameters (of some type that is not specified here), *one possible* reaction of the reservation branch is the emission of a *check_availability* message to the pickup branch. If the receiving pickup branch replies with message *available* then the reservation branch has to send an *offer* message to the requesting customer, and so on.

Assigning a meaning to sequence diagrams in the way described here corresponds to a loose semantics interpretation (cf. Section 3.3).

From a methodological point of view, scenarios may serve as a basis for complete interaction specifications; the latter may, for instance, be obtained by composing the former (cf. Section 2.2). Scenarios may also be used to derive more detailed behavior specifications, such as state transition diagrams.

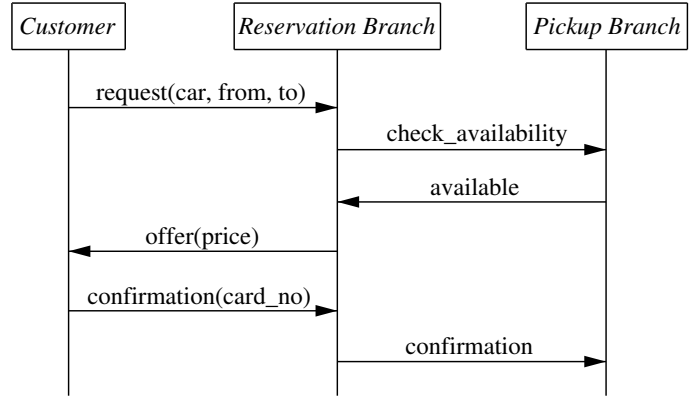

Figure 2: Exemplary Description of a Car Reservation

## 2.2 Transition to Complete Behavioral Descriptions

The previous section outlines one typical scenario that happens when a customer reserves a car in a reservation branch. The diagram in Figure 2 depicts the interactions that occur when the requested car is available, and the customer accepts the price. In order to give a more complete description of the possible interactions in a reservation branch, i.e. all possible interactions between a customer, the reservation branch, and the pickup branch, we first have to analyze whether there are other possible interaction sequences.

In the current example, we identify two additional scenarios, depicted in Figure 3 (a) and (b). Figure 3 (a) describes the scenario when a customer wishes to reserve a car that is not available for the requested period. Figure 3 (b) treats the situation where the car is available, but the customer does not accept the price.

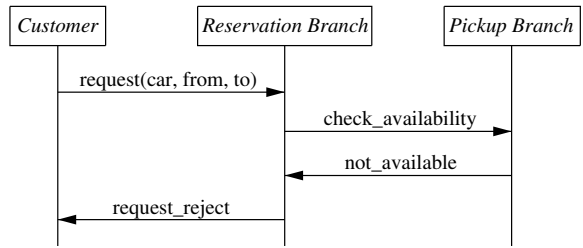
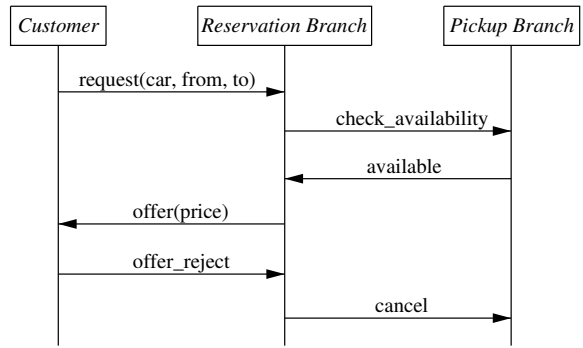

(a) (b)

Figure 3: Exemplary Descriptions of an Unsuccessful Car Reservation

Now, having specified all relevant scenarios (of this simplified example), we have to clarify the relationships between them, i.e. we have to analyze in which sequence they may occur. Afterwards, the scenarios must be composed, according to the sequences found, i.e. the particular scenarios have to be combined to more general ones. For that purpose, we use the following operators that are provided by EETs:

**Sequential composition** allows to concatenate a sequence of scenarios.

**Finite Choice** (represented by a box referencing a finite set of scenario descriptions) specifies that, when the box is to be unfolded, one element from the set has to be chosen.

**Repetition** (indicated by a line labeled with the number of iterations to the right of the EET) denotes part of a scenario description that may occur optionally or repeatedly. The vertical dimension of the indicator designates the operator's scope.

Based on the assumption that, in our example, a customer can either make (at most) one successful reservation after an arbitrary number of unsuccessful attempts we obtain the diagram depicted in Figure 4. This diagram does not describe a single scenario, but the set of all possible interaction sequences. First of all, there may be a finite number of instantiations of the box labeled with *Failed Reservation*. Each time the box is instantiated one scenario contained in the set is chosen. After that, there may be zero or one instantiations of the box labeled *Successful Reservation*.

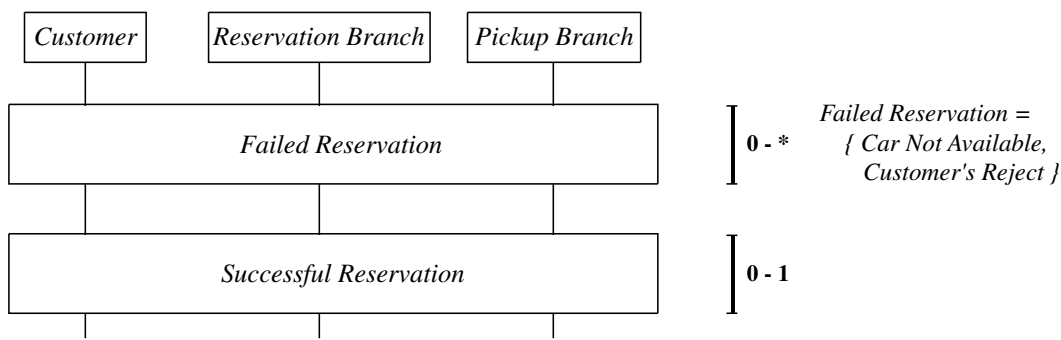

Figure 4: Complete Description of a Car Reservation

## 3 Semantics

In this section we discuss some of the issues that have to be addressed when assigning a formal semantics to sequence diagrams. To that end, in Section 3.1, we briefly mention the basic concepts (such as the communication primitives) of the mathematical model we target at. Section 3.2 contains a rough sketch of how EETs can be assigned a trace semantics. In Section 3.3 we explain why this semantics fits well into an integrated mathematical modeling technique that covers the entire system development process.

## 3.1 Basic Concepts

Most concrete (physical) systems have a particular architecture (reflected, for instance, by the system's decomposition into components and their relationships), a particular state (from the set of possible system states) and a particular behavior. As a consequence, one approach at a complete specification of such systems consists of providing several models, each being concerned with one of the above aspects. Sequence diagrams model the behavior of a system by describing the interactions among the system's components and between its components and the environment.

For the purpose of defining a semantics for a given EET we consider an *interaction* to be a message exchanged between two participating components. An EET then describes a set of *interaction sequences*. Other system properties, such as component states, are abstracted away. The only purpose of the components on this level of abstraction is to send and to receive messages. The meaning of an interaction description is an *assertion* about the possible message exchanges among the participating components. As discussed in Section 2, this assertion may describe the complete set of interaction sequences or only some constraint on the set of possible sequences.

Alternatively, one can consider *events*, i.e., particular occurrences of interactions instead of interactions and *partial orders* between events instead of interaction sequences. Moreover, one can distinguish between the event of *sending* a message and the event of *receiving* a message. This approach is used for MSCs [IT96] and allows the designer to model message overtaking.

There is a close connection between the choice of atomic actions (events or interactions) and the communication paradigm of the system modeled by sequence diagrams. We distinguish between *synchronous* and *asynchronous* communication. The former allows the receiver to block the sender, i.e., the communication takes place only when both communication partners are ready to communicate. In this case sending and receiving together constitute an atomic event. In the second case, the receiver cannot block the sender, i.e., the sender can pass a message that is processed later on by the receiver. This is usually accomplished by providing buffers between senders and receivers. In this case sending and receiving are different events.

The EETs as introduced above use synchronous communication while MSCs use asynchronous communication. Both paradigms have their merits and each of them is more natural for a particular problem setting. For example, when modeling a network protocol, it is more natural to consider asynchronous communication between the network nodes and synchronous communication inside each node. We have adopted the synchronous paradigm for EETs because we consider it to be particularly useful for modeling large parts of object-oriented business information systems (such as transaction management within database applications).

A special mechanism that is very important in an object-oriented setting is remote procedure call. Here, the messages are partitioned into operation calls, which invoke an operation on the receiving object, and returns, which represent the result of a previous call. Note that the remote procedure call shares properties with both communication paradigms mentioned above: on the one hand the receiver may not block the sender, on the other hand the sender is blocked until the receiver is willing to deliver the result. Such basic mechanisms can be dealt with in two different ways. One approach is to encode them directly into the definition of the semantics for EETs. Another one is to represent these mechanisms as constraints on a more general semantic model (cf. Section 3.3) covering other system views as well.

## 3.2 Synchronous Trace Semantics

As we have already pointed out, the semantics of an interaction description strongly depends on the underlying communication paradigm. EETs are based on synchronous message exchange. In this section, we describe their semantics informally. A more thorough semantical treatment can be found in [BHKS97, SHB96]. For the semantics of MSCs we refer the interested reader to [IT95], where it is defined in a process-algebraic setting.

The graphical constructs we consider are: empty EET, single message exchange, choice between different EETs, sequential composition of EETs, iteration (repetition) of an EET and interleaving (parallel composition) of two EETs. The intuitive graphical syntax of these operators (except interleaving) was introduced in the previous section. The interleaving operator may, for instance, be used to model concurrent reservations by different customers at the same reservation branch in the example started in Section 2. Note that the messages occuring in the graphical syntax may contain formal parameters. For the informal definition of the semantics we mimic the inductive definition presented in [BHKS97].

- An empty EET describes a single interaction sequence: the empty sequence. It is used to express the absence of communication among a set of components.

- An EET denoting a single message exchange defines a set of singleton interaction sequences, each interaction containing a message where every formal parameter is instantiated with a value of the parameter's domain.

- The operator representing the choice between two EETs has the union of the sets of interaction sequences described by each of the EETs as its semantics.

- Sequential composition of two EETs describes the set of interaction sequences obtained by collecting all possible concatenations of interaction sequences denoted by the first EET with interaction sequences denoted by the second one.

- The iteration operator applied to an EET yields the set of interaction sequences obtained by concatenating an arbitrary number of times the interaction sequences of the operand EET.

- The interleaving of two EETs describes the set of interaction sequences obtained by collecting all possible interleavings of the interaction sequences of the operand EETs.

To increase the expressive power of the description technique we may use predicates to constrain the sets of interaction sequences allowed by an EET.

One can also consider extensions of EETs by taking component states into account and by using this state to trigger interactions. Moreover, one can imagine additional composition mechanisms and several notions of refinement for EETs. We currently also examine an adequate method-call operator suitable for modeling object-oriented systems with EETs, as well as other "special purpose" operators.

## 3.3 Integration of the Semantics

One of the most important issues that have to be addressed when introducing a certain description technique is to define its relationship with existing techniques. This is especially true if the latter's scope is not orthogonal to the former's. Then, the definition and maintenance of consistency among the results of applying different notations is a prerequisite for assuring correctness in the development process.

One approach that facilitates providing an integrated semantics for different description techniques has been explored in the SysLab project. Here, syntactic diagrams (also called documents in this context) are assigned a semantics by defining a mapping that relates each diagram to a set of systems. The notion of a system is defined using an abstract mathematical model (cf. [KRB96], an extension appears in [GKR96]). The description of a system consists of various documents that describe, for instance, the properties of classes, objects, and object behavior. Every document represents a certain view on the system. One such view is the semantic model of object interaction as described in the previous sections.

The notion of consistency of interaction descriptions with other description techniques can be defined by providing mappings that extract the communication information from the latter. The result of this extraction has to be matched against the set of interaction sequences as defined by the EET document. At this point the distinction between exemplary and complete interaction descriptions comes into play. If we use EETs to represent scenarios, the matching process has to ensure the following: in the set of interaction sequences extracted from the other documents there must be at least one element that is equivalent to the semantics of the scenario. This corresponds to a loose semantic interpretation. Complete interaction descriptions, on the other hand, yield another notion of consistency: Every element of the set of extracted interaction scenarios must be contained in the semantics of the corresponding EET document. In either case an EET document may be seen as a set of proof obligations that have to be discharged to establish the consistency among the documents representing the system.

Because the semantics of documents, such as EETs, is given by the set of all systems that obey the induced properties each document can be seen as a proposition about the system to be implemented. Several documents can then be combined at the semantics level by conjoining their semantics (which corresponds to intersection on sets). Refinement and composition also have a semantic counterpart: they result in set inclusion. This means that adding information to a document always results in a smaller set of systems. For a complete integration of EETs into the system development process we have to define adequate notions for their refinement. This is an active area of work in our group.

## 4  Conclusion

In the previous sections we have both explored the application of sequence diagrams as a description technique for exemplary and complete object interactions, and outlined aspects of their formal foundation. Each of the two different semantic interpretations has its specific methodological implications.

There is ample opportunity for future research in the following areas. First, we are investigating steps towards providing developers with a methodological framework for the transition from exemplary to complete interaction descriptions. Second, we explore the combination of interaction descriptions with other description techniques, like state transition diagrams, to integrate the former into a more general design context. This requires studying adequate notions of consistency among these notations. Together with the concepts presented in this paper, these steps form the basis for integrated tools supporting developers in the design of large, object-oriented business information systems.